\documentclass[10pt,a4paper,aps,onecolumn,showpacs,amssymb,amsmath, color,floatfix]{revtex4}
\usepackage[utf8x]{inputenc}
\usepackage[english]{babel}
\usepackage[T2A]{fontenc}
\usepackage{ucs}
\usepackage{amsmath}
\usepackage{amsfonts}
\usepackage{amssymb}
\usepackage{graphicx}
\graphicspath{{EPS_figs/}{PDF_figs/}}
\usepackage{color}
\usepackage{bm}

\begin{document}

\title{Surface induced structures in nematic liquid crystal colloids}

\author{S. B. Chernyshuk}
\affiliation{Institute of Physics, NAS of Ukraine, Prospekt Nauky 46, Kyiv 03650, Ukraine}

\author{O. M. Tovkach}
\affiliation{Bogolyubov Institute for Theoretical Physics, NAS of Ukraine, Metrologichna 14-b, Kyiv 03680,Ukraine}

\author{B. I. Lev}
\affiliation{Bogolyubov Institute for Theoretical Physics, NAS of Ukraine, Metrologichna 14-b, Kyiv 03680,Ukraine}


\begin{abstract}
We predict theoretically the existence of a class of colloidal structures in nematic liquid crystal (NLC) cells, which are induced by surface patterns on the plates of the cell (like cells with UV-irradiated polyimide surfaces using micron size masks). 
These bulk structures arise from non-zero boundary conditions for the director distortions at the confining surfaces. 
In particular, we demonstrate that quadrupole spherical particles (like spheres with boojums or Saturn-ring director configurations) form a square lattice inside a planar NLC cell, which has checkerboard patterns on both its plates.

\end{abstract}

\maketitle

Liquid crystals are state of matter with an orientational ordering. Colloidal particles in such media distort the ordering and due to that interact with each other. As a result they form different one-dimensional (1D) linear \cite{Poulin_1998, Poulin_1997} and inclined \cite{Poulin_1998, Smalyukh_2005, Kuzmin_2005, Kotar_2006} chains as well as two-dimensional (2D) crystals \cite{Nazarenko_2001, Smalyukh_2004, Musevic_2006, Skarabot_2007, Skarabot_2008, Ognysta_2007}. 
Recently the authors of \cite{nych1} observed a three-dimensional (3D) colloidal crystal formed by spherical particles with the dipole configuration of the director field in their vicinity. 
The vast majority of those structures was assembled using laser tweesers manipulations with each particle. 
Because of that it is hard to achieve large sizes of the crystals in such a way. 

Besides, all those structures were obtained in NLC cells with the fixed uniform (either planar or homeotropic) director orientation at the plates. 
However, as early as in 1991, Gibbons et al. showed that the direction of the homogeneous alignment of LC molecules on specially designed polymers could be established using polarized light (a photoalignment method) \cite{Gib}. 
The photoalignment method was developed in many subsequent papers \cite{Rez1,Rez2,Sch}. 
Combining the photoalignment method with using of different masks one can obtain different patterns of the nonuniform director orientation at the  cell plates. For instance, LC cells with an aligned square lattice can be produced from polyimide film by a UV exposure technique. 
The alignment direction is determined by the polarization of the UV light and the alignment patterns can be controlled using a mask with designed opening patterns in front of the substrate \cite{China}.

In this Letter we consider theoretically colloidal particles suspended in a NLC cell with similar square lattices at both substrates. 
We demonstrate that spherical particles with the Saturn-ring or boojums director configuration should form a 2D square lattice (with a period $\sqrt{2}l$) in a planar nematic cell with checkerboard pattern (of a side $l$) at both substrates.

Consider colloidal particles in a NLC cell with some patterns at both plates. 
This means that small director deviations $n_{\mu}$, where $\mu=x, y$, from its ground state $\mathbf{n}_0=(0,0,1)$ might be nonvanishing at confining surfaces $\Sigma$.  Theoretical description of the axially symmetrical particles in NLC colloids is based on the effective free energy functional proposed in \cite{lupe}
\begin{equation}\label{Lubensky_FE}
F = K \sum_{\mu=x,y} \int d\mathbf{x} \left\{ \frac{(\nabla n_{\mu} \cdot \nabla n_{\mu})}{2} -4 \pi P(\textbf{x})\partial_{\mu} n_{\mu} -4 \pi C(\textbf{x}) \partial_{z} \partial_{\mu} n_{\mu}  \right\},
\end{equation}
where $K$ is the Frank elastic constant, $P(\textbf{x})$ and $C(\textbf{x})$ are elastic dipole and quadrupole moment densities.
Then $n_{\mu}$ obey the Poisson equations
\begin{equation}\label{EL_eq}
\Delta n_{\mu} = 4 \pi \left[ \partial_{\mu} P(\textbf{x}) - \partial_{z} \partial_{\mu} C(\textbf{x}) \right].
\end{equation}
Previous studies were focused on the case of $n_{\mu}(\mathbf{s})=0$ for any $\mathbf{s} \in \Sigma$. 
In this Letter we assume more general boundary conditions $n_{\mu}(\mathbf{s})\neq 0$ for some $\mathbf{s} \in \Sigma$. 

Under such circumstances, one can find solutions to \eqref{EL_eq} via appropriate Green's functions $G_{\mu}(\textbf{x}, \textbf{x}^{\prime})$
\begin{equation}\label{EL_sol}
n_{\mu}(\mathbf{x}) = \int_{V} d\mathbf{x}^{\prime} G_{\mu}(\textbf{x}, \textbf{x}^{\prime}) \left[ -\partial_{\mu}^{\prime} P(\textbf{x}^{\prime}) + \partial_{z}^{\prime} \partial_{\mu}^{\prime} C(\textbf{x}^{\prime}) \right] -\frac{1}{4\pi}\oint_{\Sigma}d\mathbf{s}^{\prime} n_{\mu}(\mathbf{s}^{\prime})\frac{\partial G_{\mu}}{\partial \mathbf{n}^{\prime}}(\textbf{x}, \textbf{s}^{\prime}), 
\end{equation}
where $\Delta G_{\mu}(\textbf{x}, \textbf{x}^{\prime}) = -4 \pi \delta(\textbf{x} - \textbf{x}^{\prime})$ for any $\textbf{x}, \textbf{x}^{\prime} \in V$, $G_{\mu}(\textbf{x}, \textbf{s}) =0$ for any $\textbf{s} \in \Sigma$ and $\mathbf{n}^{\prime}$ is the outer normal to $\Sigma$ \cite{Jack}.
Substituting \eqref{EL_sol} into \eqref{Lubensky_FE} and implying the superposition principle for the elastic moment densities of $N$ particles, $P(\textbf{x}) = \sum_{i=1}^{N} p_{i}\delta(\textbf{x} -\textbf{x}_{i})$ and $C(\textbf{x}) = \sum_{i=1}^{N} c_{i}\delta(\textbf{x} -\textbf{x}_{i})$ ($p_{i}$ and $c_{i}$ are dipole and quadrupole elastic moments of the $i$-th particle),  we can easily see that the free energy of the colloidal system can be written as
\begin{equation}
F = \sum_{i<j} U_{ij} +\sum_{i=1}^{N} \left[ U_i + U_i^{pattern} \right],
\end{equation}
where $U_{ij}$ is the pair interaction energy between the $i$-th and $j$-th particles and $U_i + U_i^{pattern}$ is the one-particle energy. Both $U_{ij}$ and $U_i$ arise from the bulk and zero boundary conditions $n_{\mu}(\mathbf{s})=0$ at the surfaces $\Sigma$ and were discussed earlier in \cite{we, we2}. 
The last summand $U_i^{pattern}$ originates from non-zero patterns $n_{\mu}(\mathbf{s})\neq 0$ at the surfaces $\Sigma$
\begin{equation}\label{U_s}
U_i^{pattern} = \frac{K}{2} p_i \sum_{\mu=x,y} \oint_{\Sigma} d\mathbf{s}^{\prime} n_{\mu}(\mathbf{s}^{\prime}) \partial_{\mu}\partial_{\mathbf{n}}^{\prime} G_{\mu}(\mathbf{x}_i, \mathbf{s}^{\prime}) + \frac{K}{2} c_i \sum_{\mu=x,y} \oint_{\Sigma} d\mathbf{s}^{\prime} n_{\mu}(\mathbf{s}^{\prime}) \partial_{z}\partial_{\mu}\partial_{\mathbf{n}}^{\prime} G_{\mu}(\mathbf{x}_i, \mathbf{s}^{\prime}).
\end{equation} 

\begin{figure}
\includegraphics[width=.45\textwidth]{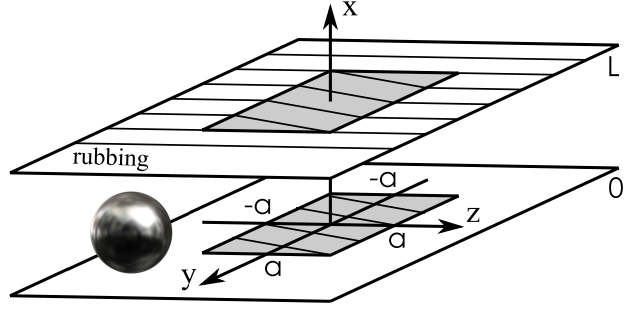}
\caption{Sketch of the planar cell with two identical squares at the surfaces $x=0$ and $x=L$.  $n_{y} > 0$ within the squares and $\textbf{n}||z$ elsewhere.}\label{Cell}
\end{figure}
\paragraph{\textbf{One square pattern at both surfaces}}

Consider first a colloidal particle suspended in a planar cell (see Fig.\ref{Cell}) with two identical squares at the $x=0$ and $x=L$ plates
\begin{equation}\label{BC}
\begin{split}
n_x &= 0,\\
n_y &= \begin{cases}
u ,\,\, &\left|z\right|\leq a\, \text{and}\, \left|y\right|\leq a\\
0,\,\, &\left|z\right| > a\, \text{and}\, \left|y\right| > a
\end{cases},
\end{split}
\end{equation}
where $u$ is a small positive constant. The Green function for a planar cell is well known \cite{we, we2}
\begin{equation}\label{GF}
G_x=G_y=\frac{4}{L}\sum_{n=1}^{\infty}\sum_{m=-\infty}^{\infty}e^{i m (\varphi-\varphi^{\prime})}\sin\frac{n \pi x}{L}\sin\frac{n \pi x^{\prime}}{L} I_{m}(\lambda_{n} \rho_{<})K_{m}(\lambda_{n}\rho_{>}),
\end{equation}
here $I_{m}$, $K_{m}$ are modified Bessel functions, $\tan\varphi=\frac{y}{z}$, $\tan\varphi^{\prime}=\frac{y^{\prime}}{z^{\prime}}$, $\rho_{>/<}$ is larger/smaller of  $\rho=\sqrt{z^2+y^2}$ and $\rho^{\prime}=\sqrt{z^{\prime 2}+y^{\prime 2}}$.
Say for simplicity that the particle is of spherical shape and the director field in its vicinity has the quadrupole symmetry.
For such a director distribution $p=0$ and $c \sim r_0^3$, where $r_0$ is the particle's radius.
In analogy with classical electrostatics $U_i$ can be treated as the energy of the interaction between the particle and all its mirror images.
Due to the symmetry of the problem $U_i=U_{i}(x)$ depends only on the $x$ coordinate of the particle and reaches its minimum in the middle of the cell $x=L/2$.
On the same grounds the other part of the one-particle energy, $U_i^{square}$, as a function of $x$ is minimal at $x=L/2$ as well. 
But $U_i^{square}$ depends also on the particle's position in the $yz$ plane as it follows from \eqref{U_s} and \eqref{GF}\begin{multline}
U_i^{square}(\frac{L}{2},y, z) = -\frac{4 \pi K c u}{L^2} \sum_{i=1}^{\infty}n \sin\frac{n \pi}{2} \left[ K_0\left(\frac{n\pi}{L}\sqrt{(y - a)^2+(z - a)^2}\right) +K_0\left(\frac{n\pi}{L}\sqrt{(y + a)^2+(z + a)^2}\right)\right. \\
\left. -K_0\left(\frac{n\pi}{L}\sqrt{(y + a)^2+(z - a)^2}\right) -K_0\left(\frac{n\pi}{L}\sqrt{(y - a)^2+(z + a)^2}\right) \right].
\end{multline}

\begin{figure}
\includegraphics[width=.9\textwidth]{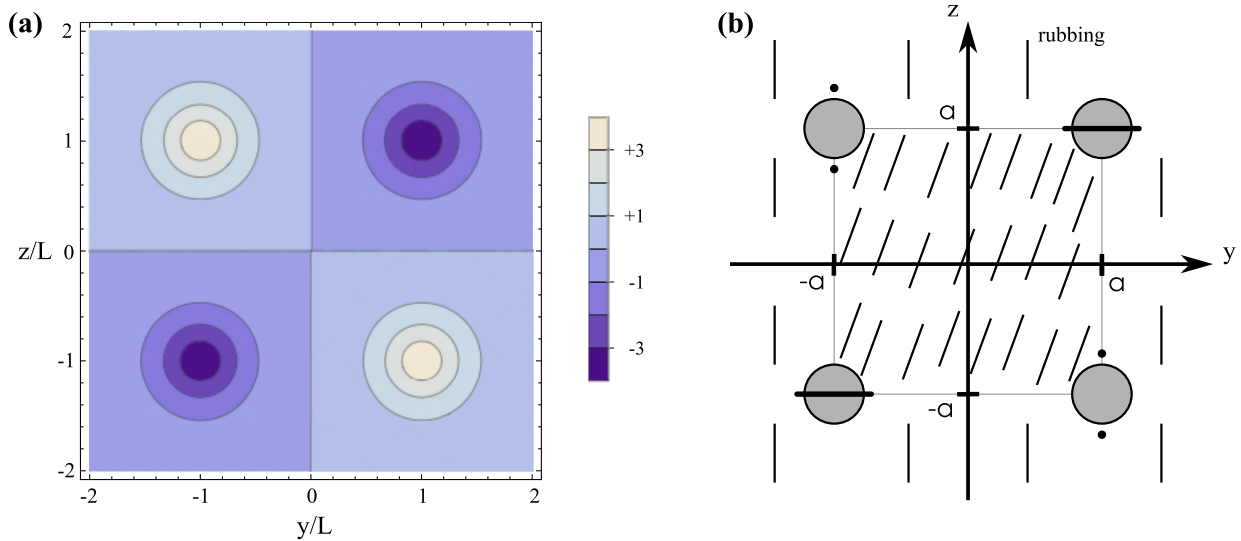}
\caption{Color online. \textbf{(a)} Contour plot of the dimensionless one-particle energy $(2K|c|u)^{-1} L^2 U_i^{square}(L/2,y, z)$ as a function of the particle's position in the $yz$ plane. Here $a=L$.
\textbf{(b)} Sketch of the equilibrium positions of the Saturn-ring and boojums particles in the cell with square pattern. All the particles are located in the middle of the cell, $x=L/2$. The segments depict the local rubbing directions.}
\label{Contour}
\end{figure}

\begin{figure}[h!]
\includegraphics[width=.45\textwidth]{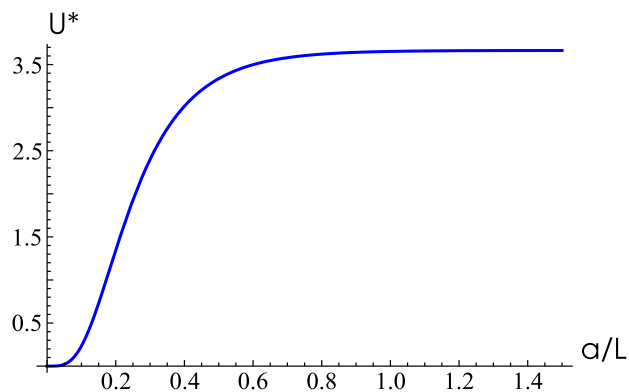}
\caption{Dimensionless depth of the potential well in the one-particle energy landscape as a function of the halflength $a$ of the one square pattern, $U^{*}=(2 K|c| u)^{-1} L^2 U_{i}^{square}(L/2,a, a)$. Maximum value of $U^{*}$ is equal to $4\beta(2) \approx 3.67$ with $\beta(r)$ being the Dirichlet beta function.}\label{Depth}
\end{figure}

Figure \ref{Contour} clearly shows that due to this dependence a particle with a positive quadrupole moment $c>0$ (like a sphere accompanied by the Saturn-ring topological defect) will be located at one of the points $(L/2, \pm a, \pm a)$.
At the same time negative quadrupoles $c<0$ (like spheres with the boojums) should be localized at the points $(L/2, \pm a, \mp a)$.
Since the Bessel function $K_0(r)$ decays exponentially at $r\gg 1$, the extremum values of the one-particle energy hardly depends on large $a$ (see Fig.\ref{Depth}).
Thus, when $a\gg L$, the maximum depth of the potential well (the height of the potential barrier simultaneously) reads as $\left|U_{i, min}^{square}\right| = U_{i, max}^{square} = \frac{4 K |c| u}{L^2} \lim_{r \to 0} \sum_{i=1}^{\infty} n \pi \sin\frac{n \pi}{2} K_0\left(\frac{n \pi r}{L} \right) = \frac{8 K |c| u}{L^2} \beta(2)$, where $\beta(r)$ denotes the Dirichlet beta function.

Here we should remark that a significant role in the formation of the usual colloidal structures belongs to higher-order elastic terms \cite{Stas, we_octupole}. They change free energy \eqref{Lubensky_FE} to be $F=K\int d^{3}x\left\{\frac{(\nabla n_{\mu})^{2}}{2}-4\pi\sum^{N}_{l=1}A_{l}(\textbf{x})\partial_{\mu}\partial_{z}^{l-1}n_{\mu} \right\}$ with $A_{l}(\textbf{x})=a_{l}\delta(\textbf{x})$ and $a_{l}$ being the higher-order moments ($a_{1}=p,a_{2}=c$). For the quadrupole particle all odd $a_{l}=0$, all even $a_{l}=b_{l}r_{0}^{l+1}$. For instance $b_2 \approx -0.3$, $b_4 \approx -0.02$, $b_6 \approx -0.0002$ for the boojums configuration \cite{Stas}. They alter $U_i^{square}(L/2,y, z)$ and make the maximum depth of the potential well to be
\begin{equation}
\left|U_{i,min}^{square}\right| = \frac{2 K u r_0^3}{L^2}\left\{ 4 \left|b_2\right| \beta(2) -\frac{3 \left|b_4\right|r_0^2}{16 L^2}\left[\zeta\left(4, \frac{1}{4}\right) -\zeta\left(4, \frac{3}{4}\right)\right] + \frac{45 \left|b_6\right|r_0^4}{64 L^4}\left[\zeta\left(6, \frac{1}{4}\right) -\zeta\left(6, \frac{3}{4}\right)\right]\right\}
\end{equation}  
where $\zeta(r, s)$ is the generalized Riemann zeta function. Thus, the contribution of the high-order elastic terms to the one-particle energy does not exceed $20\%$.

\begin{figure}
\includegraphics[width=.45\textwidth]{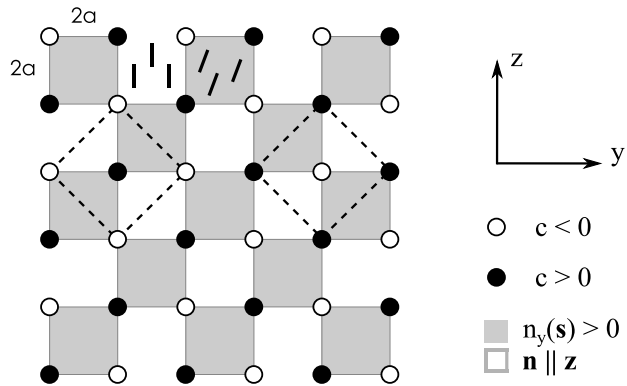}
\caption{Sketch of two possible square quadrupole lattices (black or white) in the planar nematic cell with checkerboard pattern at both surfaces. The dashed squares depict the unit cells of the lattices. All particles are located in the middle of the cell, $x=L/2$.}\label{Lattice}
\end{figure}

\paragraph{\textbf{Checkerboard pattern at both surfaces}}

\begin{figure}
\includegraphics[width=.45\textwidth]{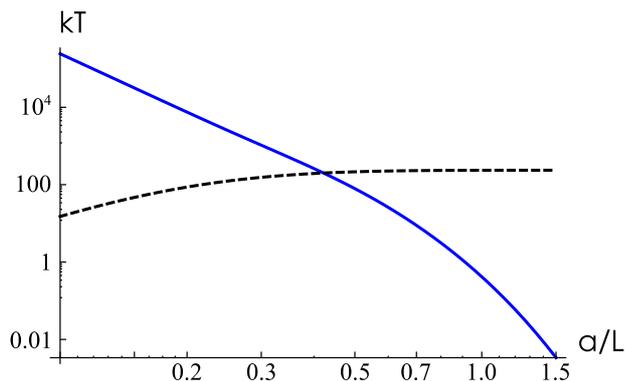}
\caption{Color online. Magnitudes of the doubled energy $2\left|U_{QQ}\right|$ of the interaction between two elastic quadrupoles located at neighboring vertices of the unit cell  (solid line) and the one-particle energy $\left|U_{i}^{ch.board}\right|$ (dashed line) as functions of $a/L$.
Here $r_0 = 2\, \mu$m, $L=6\, \mu$m, $K=10$ pN, $u=0.1$ and $c = 0.3r_0^3$. 
Under these conditions $\left|U_{i}^{ch.board}\right| \gtrsim 20 \left|U_{QQ}\right|$ for $a \gtrsim 0.65L$.}\label{Comparison}
\end{figure}

Suppose now that there are $N$ identical quadrupole particles and the boundary conditions \eqref{BC} are replicated periodically with the period $4a$ in both directions $y$ and $z$.
Obviously, this doubles the values of the one-particle energy at the nodes of the lattice.
Let us assume that each the potential well  $U_{i}^{ch.board} = - 2 \left|U_{i}^{square}(L/2,a, a)\right|$ is occupied by one of the particles.
Then the unit cell of such a structure is a square with a side of length $a\sqrt{8}$ (see Fig.\ref{Lattice}).
The cell's energy is the sum of $U_{i}^{ch.board}$ and $2 U_{QQ}$ with
\begin{equation}\label{U_QQ}
U_{QQ} = -4\pi K c_i c_j \partial_{z_i}\partial_{z_j} \left[\partial_{x_i}\partial_{x_j} G_x(\mathbf{x}_i, \mathbf{x}_j) +  \partial_{y_i}\partial_{y_j} G_y(\mathbf{x}_i, \mathbf{x}_j) \right]
\end{equation}
being  the energy of the quadrupole-quadrupole interaction between the particles located at neighboring vertices of the unit cell (an explicit expression for $U_{QQ}$ was derived in \cite{we2} but it is too cumbersome to be presented here).
The interaction between opposite vertices can be omitted because of its smallness.
Thus, such a surface induced colloidal structure will be stable if $\left|U_{i}^{ch.board}\right|$ prevails both thermal fluctuations and $2 \left|U_{QQ}\right|$. Then comparing the magnitudes of the one-particle and the doubled interparticle energies one can conclude that $\left|U_{i}^{ch.board}\right|$ is at least 10 times greater than $2\left|U_{QQ}\right|$ for $a \gtrsim  0.65L$ (see Fig.\ref{Comparison}).
At these side lengths the depth of the potential well $\left|U_{i}^{ch.board}\right| \approx 2\left|U_{i,min}^{square}\right| \approx 240 kT$ is large enough for the structure to be experimentally observable. Moreover, if at the same time the concentration of the particles is about $(8La^2)^{-1}$, i.e. a particle per a unit cell, we may expect their self-organization into the lattice described above. Note also that in a usual thin planar cell the elastic quadrupoles arrange in a close-packed lattice with a parallelogram unit cell \cite{Musevic_2006}.

In conclusion, we have shown that patterning of the surfaces in nematic cells can induce colloidal structures inside of them. Characteristics of such structures are governed by the one-particle energies and may be substantially different from those of the usual colloidal structures, which are formed by interparticle interactions. In particular, we predict theoretically, that spherical particles with the Saturn-ring or boojums director configuration will form a 2D square lattice in a planar cell with checkerboard pattern at both substrates.

\end{document}